# Obtaining phase diagram and thermodynamic quantities of bulk systems from the densities of trapped gases

Tin-Lun Ho[1] and Qi Zhou[1],

*1. Department of Physics, The Ohio State University, Columbus, OH 43210*

**At present, many laboratories are performing experiments to simulate theoretical models of strongly correlated systems using cold atoms in optical lattices, a program referred to as "Quantum Simulation"[1,2,3,4,5,6,7,8]. It is hoped that these experiments will shed light on some long standing problems in condensed matter physics. The goal of Quantum Simulation is to obtain information of homogenous bulk systems. However, experiments are performed in confining traps. The non-uniformity of the trapping potential inevitably introduces different phases in the sample, making it difficult to deduce the properties of a single bulk phase. So far, there are no algorithms to use the experimental data to map out phase diagrams and important thermodynamic quantities. Here, we present an algorithm to achieve this goal. Apart from phase diagram, it also maps out entropy density, superfluid density of superfluids, and staggered magnetizations of anti-ferromagnets. Our scheme is exact within local density approximation.**

To deduce the bulk properties of homogenous systems from the observed properties of non-uniform systems, local density approximation (LDA) naturally comes to mind. This approximation assigns the properties of a non-uniform system at a given point their bulk values with an effective local chemical potential. To the extent that LDA is valid, determining the bulk thermodynamic quantities as functions of chemical potential amounts to determining their spatial dependences in confining traps. In current experiments on ultra cold atomic gases, the detection closest to a local probe is the measurement of column density. Here, we introduce algorithms to determine the quantities mentioned in the opening using column densities.

Our first step is to use the surface density a thermometer. Within LDA, the density (both for atoms in a single trap or in an optical lattice) is $n(\vec{x}) = n(\mu(\vec{x}),T)$ where $n(\mu,T)$ is the density of a homogenous system with temperature $T$ and chemical potential $\mu$, $\mu(\vec{x}) = \mu - V(\vec{x})$, and $V(\vec{x}) = \frac{1}{2} M \sum_{i=x,y,z} \omega_i^2 x_i^2$ is a harmonic trapping potential with frequency $\omega$. Near the surface, the density is sufficiently low, so one can perform a systematic fugacity expansion. The surface density of a quantum gas is then

$$n(\vec{x}) = \alpha e^{(\mu-V(\vec{x}))/T} / \lambda^3 \qquad (1)$$

where $\alpha = 1$ for a single trap, and $\alpha = \left[I_0(2t/T)d\right]^3$ for gases in optical lattices, and $I_0(x)$ is the Bessel function of the first kind (See supplementary materials). The corresponding column density $\tilde{n}(x,y) = \int n(x,y,z)dz$ is, (with $\vec{r} = (x,y)$) is

$$\tilde{n}(x,y) = \alpha \left(\frac{T}{\hbar\omega}\right) \frac{e^{(\mu-V(\vec{r}))/T}}{\lambda^2} \qquad . \qquad (2)$$

Eq.(2) has been widely used to determine $\mu$ and $T$ of quantum gases in single traps but not for gases in optical lattices, as the surface density is very low in such cases. The lack of accurate thermometry in optical lattices has been the bottleneck for extracting information from current experiments. For example, it has prevented mapping out the phase diagram of Bose Hubbard Model at finite temperature despite many years of studies. It has also aroused the concern on the heating effects in current optical lattice experiments[9,10,11,12]. To make use of the asymptotic forms in Eq.(1) and (2), we need imaging resolutions comparable to a lattice spacing (~0.5 micron), which has not been achieved in most experiments. Very recently, however, Ott's group[13] has succeeded in imaging the density of a 3D quantum gas using a focused electron beam with extremely high resolution (0.15 micron). This exciting development shows that the capability to determine of $\mu$ and $T$ accurately using surface density is already in place.

With $\mu$ and $T$ determined from the surface density, one readily obtains the equation of state $n(v,T)$ by identifying it with $n(\vec{x})$, where $V(\vec{x}) = \mu - v$. However, in current experiments, only column density, $\tilde{n}(x,y) = \int n(x,y,z)dz$, is measured. To obtain the density $n(\vec{x})$, one can use a method of Erich Mueller which shows the pressure $P$ along the x-axis is given by integrating the column density along $y$,

$$P(x,0,0) = \frac{M\omega_y \omega_z}{2\pi} \tilde{\tilde{n}}(x), \qquad (3)$$

where $\tilde{\tilde{n}}(x) = \int \tilde{n}(x,y)dy$, and $P = V^{-1}T\ln[Tre^{-(H-\mu N)/T}]$ for a homogenous system with volume $V$. It satisfies the well know Gibbs-Duham relation

$$dP = nd\mu + sdT. \qquad (4)$$

We then have

$$n(x,0,0) = \left(\frac{\partial P}{\partial \mu(x)}\right)_T = -\frac{1}{2\pi x}\frac{\omega_y \omega_z}{\omega_x^2}\frac{d\tilde{\tilde{n}}(x)}{dx}. \qquad (5)$$

Eq.(3) follows directly from the fact that $\tilde{\tilde{n}}(x) = \int n(\mu - \frac{1}{2}M\sum_{i=x,y,z}\omega_i^2 x_i^2, T)dydz$, and $dydz = -\frac{2\pi}{M\omega_y\omega_z}d\mu$ for given $x$. The integral is therefore proportional to $\int nd\mu = \int dP$, hence Eq.(3).

Since singularities of thermodynamic potentials show up in equation of state, phase boundaries between different phases will be identified in the density profile.

Recall that first order and continuous phase transitions correspond to discontinuities in the first and higher order derivatives of $P$. Hence, from Eq.(4), $n$ and $s$ are discontinuous across a first order phase boundary, whereas the slope of $dn/d\mu$ and $ds/dT$ are discontinuous for higher order phase boundary. The discontinuity in $n$ has been used in the recent MIT experiment to determine the first order phase boundary in spin polarized fermions near unitarity[14]. Since $\dfrac{dn(x,0,0)}{dx} \propto \dfrac{dn(\mu(x,0,0),T)}{d\mu}$, a higher order phase boundary will show up as a discontinuity of the slope from the compressibility, which can be extracted from the density profile. The presence of such discontinuity has also seen in Monte Carlo studies, (Q.Zhou et.al, to be published).

We now turn to entropy density $s(\vec{x})$, which is useful for identifying phases. For example, for a spin-1/2 fermion Hubbard model, if $s(\vec{x})$ is far below $\ln 2$ in certain regions, it is a strong evidence for spin ordering. To obtain $s = (dP/dT)_\mu$, we need to generate two slightly different configurations of $P(\vec{x})$ with different $T$ and calculate their difference at the same $\mu$. To do this, we change the trap frequency $\omega_x$ adiabatically to a slightly different value $\omega_x'$ ($\omega_x' = \omega_x + \delta\omega_x, \delta\omega_x \ll \omega_x$). Both $\mu$ and $T$ will then change to a slightly different value, say, to $\mu'$ and $T'$[9]. One can then measure the column density of the final state and construct its pressure function $P(x,0,0)$. The entropy density of the initial state along the $x$-axis is

$$s(x,0,0) = \frac{P'(x',T') - P(x,T)}{T' - T}, \qquad (6)$$

where $x$ and $x'$ are related as

$$\mu(x) \equiv \mu - \frac{1}{2}M\omega_x^2 x^2 = \mu' - \frac{1}{2}M\omega_x'^2 x'^2 \equiv \mu'(x'). \qquad (7)$$

See Figure 2.

We next consider superfluid density $n_s$, a fundamental quantity that has not been measured in cold atom experiments. It is a quantity particularly important for 2D superfluids[15,16,17], where the famous Kosterlitz-Thouless transition is reflected in a universal jump in superfluid density. Without a precise determination of $n_s$, interpretation of experimental results, be they based on quantum Monte Carlo simulations[15,17] or on features of interference pattern[16] will always be indirect, due to the inhomogeneity of the system. Here, we propose a scheme to measure the inhomogenous superfluid density in the trap. For a superfluid, we have[18,19]

$$dP = nd\mu_o + sdT - Mn_s \vec{w} \cdot d\vec{w} \qquad (8)$$

where $n_s$ is the superfluid number density, $\vec{w} = \vec{v}_s - \vec{v}_n$, $\vec{v}_s$ and $\vec{v}_n$ are the superfluid and normal fluid velocity, respectively. $\mu_o$ is the chemical potential in the $\vec{v}_n = 0$ frame. A direct consequence of Eq.(8) is that

$$\left(\frac{\partial n}{\partial w^2}\right)_{\mu_o,T} = -\frac{M}{2}\left(\frac{\partial n_s}{\partial \mu_o}\right)_{w^2,T}. \qquad (9)$$

For a potential rotating along $\hat{z}$ with frequency $\Omega$, $\vec{v}_n = \Omega \hat{z} \times \vec{x}$. If $\Omega$ is below the frequency for vortex generation, $\vec{v}_s = 0$, and $w^2 = \Omega^2 r^2$. Since $\vec{w}$ varies in space, we cannot apply the method developed for $s(\vec{r})$. Instead, one can use the following procedure. Let $n^{(i)}(\vec{x})$ be the density of a stationary system (with temperature $T$ and chemical potential $\mu^{(i)}$) in a cylindrical trap with transverse frequency $\omega_\perp^{(i)}$ and longitudinal frequency $\omega_z$. Within LDA, we have $n^{(i)}(\vec{x}) = n(\mu^{(i)}(\vec{x});T;\vec{w}=\vec{0})$, where

$$\mu^{(i)}(\vec{x}) = \mu^{(i)} - \frac{1}{2}M\omega_\perp^{(i)2}r^2 - \frac{1}{2}M\omega_z^2 z^2. \qquad (10)$$

Next we rotate this system with frequency $\Omega$ along $\hat{z}$, and adjust $\omega_\perp^{(i)}$ to $\omega_\perp^{(f)}$ so that the temperature remains at $T$. The chemical potential then becomes $\mu^{(f)}$, and the density of this final state is $n^{(f)}(\vec{x}) = n(\mu^{(f)}(\vec{x}),T,\vec{w})$, where

$$\mu^{(f)}(\vec{x}) = \mu^{(f)} - \frac{1}{2}M(\omega_{\perp}^{(f)2} - \Omega^2)r^2 - \frac{1}{2}M\omega_z^2 z^2. \tag{11}$$

For small $w^2$, we have

$$n(\mu^{(f)}(\vec{x}),T,\vec{w}) = n(\mu^{(f)}(\vec{x}),T,\vec{0}) + \left(\frac{\partial n}{\partial w^2}\right)_{\mu^{(f)},w=0} w^2. \tag{12}$$

We then write $n(\mu^{(f)}(\vec{x}),T,0) = n(\mu^{(i)}(\vec{x}^*),T,0) \equiv n^{(i)}(\vec{x}^*)$, where $\vec{x}^* = (x^*,y^*,z^*)$, $z^* \equiv z$, $\frac{1}{2}M\omega_{\perp}^{(i)2} y^{*2} \equiv \frac{1}{2}M(\omega_{\perp}^{(f)2} - \Omega^2)y^2$, and $\mu^{(i)}(x^*,0,0) \equiv \mu^{(f)}(x,0,0)$. Using Eq.(9), we have

$$\frac{n^{(f)}(\vec{x}) - n^{(i)}(\vec{x}^*)}{M\Omega^2(x^2+y^2)/2} = -\left(\frac{\partial n_s}{\partial \mu^{(f)}}\right)_{w=0}, \tag{13}$$

where $n_s(x,y,z) = n_s(\mu^{(f)}(\vec{x}),T,\vec{0})$. Integrating Eq.(13) over $z$ and $y$, and noting that $dydz = \frac{-2\pi}{M\omega_z\sqrt{\omega_{\perp}^{(f)2} - \Omega^2}} d\mu^{(f)}$ when $x$ is constant, we have

$$n_s(x,0,0) = \frac{\omega_z\sqrt{\omega_{\perp}^{(f)2} - \Omega^2}}{\pi\Omega^2} \int dy \frac{\tilde{n}^{(i)}(x^*,y^*) - \tilde{n}^{(f)}(x,y)}{(x^2+y^2)}. \tag{14}$$

Eq.(14) gives $n_s$ in terms of the column densities of the initial and final state. The above formula continue to hold for non-axisymmetric traps, (with $\omega_{\perp}^{(f)} \to \omega_y^{(f)}$). (See also Supplementary Material for the expression for the 2D case, and an alternative scheme for obtaining $n_s(\vec{x})$).

Our method can also be applied to obtain other important thermodynamic properties such as the staggered magnetization and the contact density of strongly interacting fermion gas. For the latter, see Supplementary Materials. In the current quantum simulator programs on Fermion-Hubbard Model using two component Fermions in optical lattices[20,21], the measurement of the staggered magnetization will be crucial for identifying the antiferromagnet. Consider an antiferromagnet in a cubic lattice with a staggered magnetic field, $\vec{h}(\vec{x}) = \hat{z} e^{i\pi(n_x+n_y+n_z)}\tilde{h}$, where $\vec{x} = (n_x, n_y, n_z)d$ are the lattice sites, $n_i$ are integers, $d$ is the lattice spacing, and $\tilde{h}$ is the magnitude of the staggered field. The hamiltonian for a homogenous system is $H = H_H - \sum_{\vec{x}} \tilde{m}_{op}(\vec{x})\tilde{h}$, where $H_H$ is the Hubbard hamiltonian, $\tilde{m}_{op}(\vec{x}) = \hat{z} e^{i\pi(n_x+n_y+n_z)} m_z(\vec{x})$ is the staggered magnetization operator, and $\vec{m}(\vec{x})$ is the spin operator at $\vec{x}$. Antiferromagnetism corresponds to $\tilde{m} = \langle \tilde{m}_{op} \rangle \neq 0$ as $\tilde{h} \to 0$. It is straightforward to show that

$$dP = nd\mu + sdT + \tilde{m}d\tilde{h}. \qquad (15)$$

The staggered field $\vec{h}(\vec{x})$ has been produced recently[22]. To reduce spontaneous emission and hence heating, one can use a low intensity laser and hence a weak field $\tilde{h}$. Note that even a weak field can produce large changes in density in the spatial region close to anti-ferromagnetic phase boundary, where bulk spin susceptibility $d\tilde{m}/d\tilde{h}$ diverges. So, measuring the responses to $\tilde{h}$ can locate the phase boundary.

Since $\tilde{m} = (\partial P / \partial \tilde{h})_{\mu,T}$, we need to generate two configurations of $P$ with different $\tilde{h}$ while fixing $\mu$ and $T$. We begin with an initial state with $\tilde{h} = 0$, determine its $\mu$ and $T$ and pressure $P(x,0,0)$ as discussed before. We then turn on a weak $\tilde{h}$ adiabatically. At the same time, we adjust $\omega$ to a new value $\omega'$ so that temperature of the final state remains fixed at $T$, while the chemical is changed to $\tilde{\mu}'$. We then construct the pressure $P'(x,0,0)$ of the final state. By noting that for any point $(x',0,0)$ in the final state, one find a corresponding point $(x,0,0)$ in the initial state such that their effective chemical are identical, $x$, i.e. $\mu(x,0,0) \equiv \mu - \frac{1}{2}M\omega^2 x^2 = \mu' - \frac{1}{2}M\omega'^2 x'^2 \equiv \mu'(x',0,0)$, we have

$$\tilde{m}(x,0,0) = \frac{P'(\mu'(x',0,0),T,\tilde{h}) - P(\mu(x,0,0),T,0)}{\tilde{h}}. \qquad (16)$$

In summary, we have pointed out a scheme to map out the bulk properties of homogenous systems using solely the density profile of a trapped gas. The method is exact within local density approximation. Our scheme requires imaging resolution comparable to a lattice spacing, a condition well satisfied by an exciting experimental advance[13]. Our method allows one to determine many properties that have so far eluded measurements. We hope this work will encourage the community to develop high precision measurements to study the many-body physics of degenerate quantum gases.

This work is supported by NSF Grants DMR0705989, PHY05555576, and by DARPA under the Army Research Office Grant Nos. W911NF-07-1-0464, W911NF0710576.

Correspondence and requests for materials should be addressed to T.L Ho. (e-mail: ho@mps.ohio-state.edu) and Q. Zhou (email: qzhou@mps.ohio-state.edu).

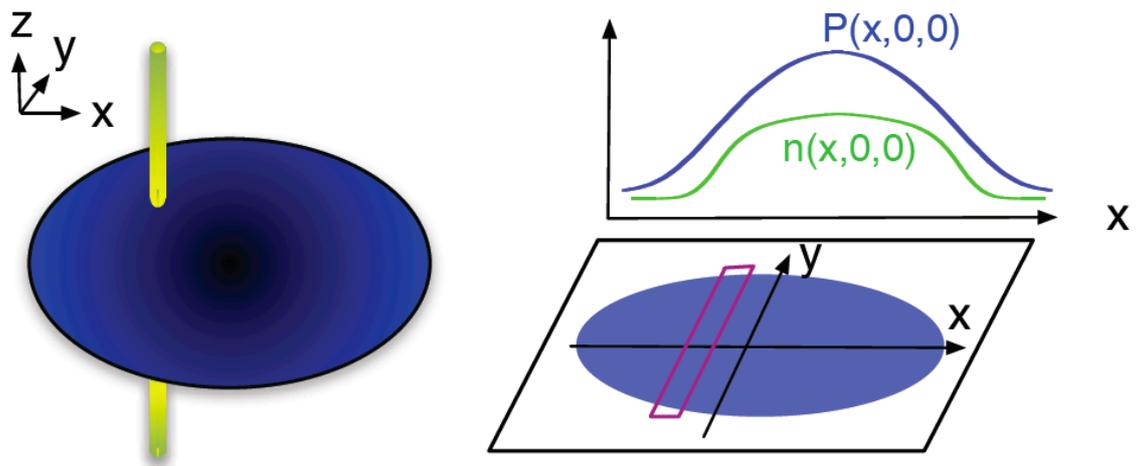

Figure 1. An illustration of the method to construct pressure from column density. The three dimensional quantum gas is represented by a blue ellipse. The green rod represents the column density collected in the experiment. By integrating the column density along y-direction, as shown in the purple box, one obtains $\tilde{n}(x)$, and hence $P(x,0,0)$ from eq.(3). The density $n(x,0,0)$ can be obtained from by differentiating $P(x,0,0)$ as in Eq.(5).

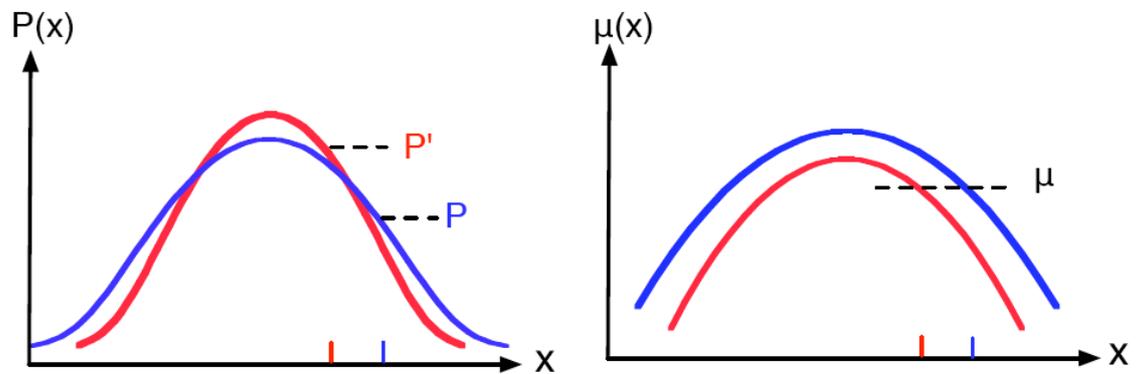

Figure 2. An illustration of the scheme for determining entropy density $s(x)$. The pressure curve $P(x,0,0)$ and effective chemical potential $\mu(x)$ of the initial state with temperature $T$ are shown in blue. The corresponding quantities of the final state are shown in red. The final equilibrium state is generated from the initial state by changing the trap frequency from $\omega$ to $\omega'$ adiabatically. To find $s(x)$,

we find the position $x'$ related to $x$ with identical effective local chemical potential by equating $\mu(x,0,0) = \mu'(x',0,0)$. The pressures at x and x' are denoted as P and P' in the figure. $s(x)$ is given Eq.(6), which is $(P'-P)/(T'-T)$.

Supplementary Material:

In the following, we shall present three new results. They are **(I)** To map out the "contact density" in strongly interacting Fermi gases. **(II)** Mapping out the superfluid density for 2D superfluids. **(III)** An alternative way to map out superfluid density. Results **(I)** and **(II)** further illustrate the range of applications of our algorithm. In result **(III)**, we show an alternative scheme to measure superfluid density which has one less step than the one mentioned in the text. This scheme will also allow on to obtain superluid density profile even when the system has unequal spin population. We shall also compare the merits of this method with that in the text. A short Appendix is also included to explain Eq.(1).

**(I) Mapping out the contact density:** Recently, it has come to light that many properties of Fermi gases with large scattering length are characterized by the so-called "contact" density, defined through the relation $[dE/d(1/a)]_S = \mathcal{V}c$, where $a$ is the scattering length[1], and the derivative is performed at constant entropy. It is then straightforward to show that

$$dP = nd\mu + sdT + cd(1/a). \tag{1}$$

Hence we have $c = \left[\partial P/\partial(\frac{1}{a})\right]_{\mu,T}$. We can then repeat the procedure in Section (**D**) with different configuration generated with $\tilde{h}$ changed to $1/a$. We then have

$$c(0,0,z) = \frac{P'(\mu'(0,0,z'),T,1/a') - P(\mu(0,0,z),T,1/a)}{1/a' - 1/a}. \tag{2}$$

**(II) Mapping out superfluid density for the 2D case:** The study of superfluid density is particularly important for the 2D case, as the superfluid density is expected to undergo a discontinuous jump at the Kosterlitz-Thouless transition. In 2D, Eq.(8) and (9) continue to hold. The initial density and the final density are now

$$n^{(i)}(\mu^{(i)}(\mathbf{r});T;\mathbf{w}=\mathbf{0}), \quad n^{(f)}(\mu^{(f)}(\mathbf{r});T;\mathbf{w}), \tag{3}$$

$$\mu^{(i)}(\mathbf{r}) = \mu^{(i)} - \frac{1}{2}M\omega_\perp^{(i)2}r^2, \quad \mu^{(f)}(\mathbf{r}) = \mu^{(f)} - \frac{1}{2}M(\omega_\perp^{(f)2} - \Omega)^2)r^2 \tag{4}$$

where $\mathbf{r} = (x, y)$. For small $\Omega$, Eq.(12) applies. We then have

$$\frac{n^{(f)}(x,y) - n^{(i)}(x^*, y^*)}{M\Omega^2(x^2+y^2)/2} = -\left(\frac{\partial n_s}{\partial \mu^{(f)}}\right)_{w=0} \qquad (5)$$

where $(x^*, y^*)$ are defined as

$$\mu^{(i)}(x^*, 0) = \mu^{(f)}(x, 0), \quad \frac{1}{2}M\omega_\perp^{(i)2}y^{*2} = \frac{1}{2}M(\omega_\perp^{(f)2} - \Omega^2)y^2. \qquad (6)$$

Unlike the 3D case, to obtain $n_s$, we multiply Eq.(5) by $ydy$ and integrate over $y$. Noting that for constant $x$,

$$-ydy = \frac{d\mu^{(f)}(\mathbf{r})}{M(\omega^2 - \Omega^2)}, \qquad (7)$$

and $n_s(x, 0) = n_s(\mu^{(f)}(x, 0); T; \mathbf{0})$, we have

$$n_s(x, 0) = \frac{2(\omega_\perp^{(f)2} - \Omega^2)}{\Omega^2} \int_{-\infty}^{\infty} \frac{n^{(i)}(x^*, y^*) - n^{(f)}(x, y)}{x^2 + y^2} ydy. \qquad (8)$$

Eq.(8) gives the superfluid density in terms of the densities at different rotation rates.

**(III) An alternative way to obtain superfluid density from the density profile:** Here is a different method to obatin $n_s$ from density profile. Since in equilibrium,

$$\nabla P = n\nabla\mu_o - n_s \nabla(M\Omega^2 r^2/2) \qquad (9)$$

where $\mathbf{r} = (x, y)$, and the rotation is along $\hat{\mathbf{z}}$. Suppose we can find the entire function $P(\mathbf{x})$ for all $\mathbf{x}$, Eq.(9) then implies that

$$n_s(\mathbf{x}) = -\frac{P(\mathbf{x}') - P(\mathbf{x})}{M\Omega^2(\mathbf{r}'^2 - \mathbf{r}^2)/2} \qquad (10)$$

where $\mathbf{x}'$ is a point close to $\mathbf{x}$ such that $\mu_o(\mathbf{x}) = \mu_o(\mathbf{x}')$, and $\mathbf{x} = (\mathbf{r}, z) = (x, y, z)$. Since $\mu_o(\mathbf{x}) = \mu_o - \frac{1}{2}(\omega_\perp^2 - \Omega^2)r^2 - \frac{1}{2}M\omega_z^2 z^2$, where $\mu_o$ is the chemical potential at the center of the trap, we have

$$(\omega_\perp^2 - \Omega^2)(r'^2 - r^2) = \omega_z^2(z^2 - z'^2). \qquad (11)$$

The question is how to construct the entire function $P(\mathbf{x})$. In general, this is difficult to do. However, in the case of cylindrical trap, one can perform the Abel transform to obtain the full density $n(\mathbf{x})$ from the column density $\tilde{n}(x,y) = \int n(x,y,z) dz$. Once $n(\mathbf{x})$ is obtained, we can make use of the equilibrium relation

$$\frac{\partial P(\mathbf{x})}{\partial z} = -n(\mathbf{x}) M \omega_z^2 z \tag{12}$$

which implies

$$P(x,y,z) = P(x,y,\tilde{z}) - \int_{\tilde{z}}^{z} dz' n(x,y,z')(M\omega_z^2 z') \tag{13}$$

for any $\tilde{z}$. Now, if we choose for any $(x,y)$ a $\tilde{z}$ such that the point $(x,y,\tilde{z})$ is at the surface of the cloud, then $P(x,y,\tilde{z})$ is the pressure at the surface, which is given with high accuracy by the expression pressure of an ideal gas. We then have

$$P(x,y,z) = P_{ideal}(x,y,\tilde{z}) - \int_{\tilde{z}}^{z} dz' n(x,y,z')(M\omega_z^2 z'). \tag{14}$$

Alternatively, we have

$$P(x,y,z) = \int_{-\infty}^{z} dz' n(x,y,z')(M\omega_z^2 z'). \tag{15}$$

In the case where the system has unequal spin population, we have

$$\begin{aligned} dP =\ & n_1 d\mu_1 + n_2 d\mu_2 + s dT - n_s d(M\mathbf{w}^2/2) & (16) \\ =\ & n d\mu + m dh + s dT - n_s d(M\mathbf{w}^2/2) & (17) \end{aligned}$$

where $n = n_1 + n_2$, and $\mu = (\mu_1 + \mu_2)/2$; $m = n_1 - n_2$, and $\mu = (\mu_1 - \mu_2)/2$. Within LDA, $\mu_i \to \mu_i(\mathbf{x}) = \mu_i - V(\mathbf{x})$, $V(\mathbf{x}) = \frac{1}{2}(\omega_\perp^2 - \Omega^2)r^2 - \frac{1}{2}M\omega_z^2 z^2$. This means $\mu \to \mu(\mathbf{x}) = \mu - V(\mathbf{x})$, whereas $h$ remains fixed. We then have

$$n_s = -\left(\frac{\partial P}{\partial (M\mathbf{w}^2/2)}\right)_{\mu,h,T}. \tag{18}$$

Comparing Eqs.(16) and (17) with Eqs.(9) and (10), we note that all discussions above for the derivation for Eq.(10) continue to hold for the case $h \neq 0$. This method therefore works for cases with unequal spin population.

The advantage of this method is that it only requires using a single density profile, whereas all the methods in the text require comparing two different density profiles. This is special for $n_s$, since the "field" the couples to it is $w^2 = \Omega^2 r^2$, which is non uniform in space. This allows one to compare pressures at different points where $w^2$ are different. This method does not work for staggered magnetization and contact density since in both cases, $\tilde{h}$ and $1/a$ are uniform in space.

The advantage of the scheme in the text, however, is that it works for harmonic traps that do not have any particular symmetry. (Although our discussions in the text is for axial symmetric traps, it is easy to see can be generalized easily to non-axisymmetric one, thought the formula for $n_s$ will be slightly different.) The use of non-axis symmetric trap is important if one needs to impart angular momentum into the system efficiently.

**Appendix**: Derivation of $\alpha$ in Eq.(1): In the lowest order in fugacity expansion in a lattice, we have $n = d^{-3} e^{\mu/T} \sum_k e^{-\epsilon_k/T}$, where the $k$-sum is over the first Brilloin Zone. Since $\epsilon_k = -2t \sum_{i=x,y,z} \cos k_i d$, where $t$ is the tunneling inktegral, we have $\alpha = [I_o(2t/d)]^3$ in Eq.(1).